# Spatially resolved quantum magnetometry and stray-field reconstruction of permalloy microdisks using boron-vacancy centers in hexagonal boron nitride

*Peiting Wen*[1,2], *Shuyu Wen*[1*], *Jiang Qu*[3], *Zeling Xiong*[1,2], *Katrin Schultheiss*[1], *Slawomir Prucnal*[1], *Artur Erbe*[1,2], *Kenji Watanabe*[4], *Takashi Taniguchi*[4], *Jürgen Lindner*[1], *Jürgen Fassbender*[1,2], *Manfred Helm*[1, 2], *Shengqiang Zhou*[1], *Ruslan Salikhov*[1], *Helmut Schultheiss*[1], *Yonder Berencén*[1**]

[1]Helmholtz-Zentrum Dresden-Rossendorf, Institute of Ion Beam Physics and Materials Research, 01328 Dresden, Germany

[2]Technische Universität Dresden, 01062 Dresden, Germany

[3]Leibniz Institute for Solid State and Materials Research, 01069 Dresden, Germany

[4]National Institute for Materials Science, 1-1 Namiki, Tsukuba 305-0044, Japan

Corresponding author: s.wen@hzdr.de[*]; y.berencen@hzdr.de[**]

## Abstract

Transferable hexagonal boron nitride (hBN) hosting negatively charged boron-vacancy ($V_B^-$) spin defects offers a versatile platform for integrated quantum magnetometry, yet quantitative imaging of magnetic microstructures remains challenging. Here, we integrate a transferred hBN flake with a 4 μm-diameter permalloy (Py = $Ni_{81}Fe_{19}$) microdisk and perform spatially resolved optically detected magnetic resonance measurements at room temperature. An applied in-plane magnetic field distorts the vortex-state magnetization, generating edge-localized magnetic surface charges and pronounced stray-field signatures at opposite disk edges. By referencing each pixel to its local zero-field splitting and correcting for a residual out-of-plane bias field, we quantitatively reconstruct the out-of-plane stray-field distribution, revealing peak fields of approximately 11.2 mT. An edge-charge model reproduces the spatial distribution and amplitude of the reconstructed field, linking the ODMR response to the field-driven evolution of the vortex state. These results establish transferred hBN $V_B^-$ sensors for quantitative magnetometry of magnetic microstructures.

## Introduction

Optically addressable spin defects in wide-bandgap semiconductors have emerged as powerful quantum sensors for nanoscale magnetometry, enabling non-invasive magnetic-field measurements under ambient conditions[1–4]. Among these systems, the nitrogen-vacancy (NV) center in diamond remains the benchmark for room-temperature quantum sensing because of its efficient optical initialization and readout, long spin coherence times, and mature quantum-control protocols[5,6]. However, the three-dimensional nature of diamond and the reliance on bulk crystals or scanning-probe geometries can limit sensor-to-sample proximity, device integration, and scalable fabrication, motivating alternative material platforms with greater integration flexibility[7–9].

Hexagonal boron nitride (hBN), a layered van der Waals material, has recently emerged as one of the most promising alternatives. In particular, the negatively charged boron-vacancy ($V_B^-$) center possesses a spin-triplet ground state with optically addressable spin transitions that remain operational under ambient conditions, making it an attractive defect for quantum sensing applications[2,10,11]. Beyond its favorable spin properties, hBN offers several unique advantages over conventional bulk hosts. Its 2D nature minimizes the sensor-to-sample separation, thereby enhancing sensitivity to near-surface magnetic fields. Furthermore, hBN flakes can be transferred onto a wide variety of materials through van der Waals assembly without the constraints associated with lattice matching or complex fabrication processes, facilitating their integration into hybrid quantum devices, functional heterostructures, and emerging spintronic platforms[10–12].

Recent studies have demonstrated the capability of $V_B^-$ centers in hBN for magnetic field sensing, microwave imaging, and quantum sensing of spin wave excitations[10,13–17]. In particular, integration with magnetic materials such as yttrium iron garnet ($Y_3Fe_5O_{12}$, YIG) has enabled optical detection of magnon excitations, underscoring the potential of hBN-based quantum sensors for hybrid quantum-magnetic systems[14,15]. However, quantitative magnetometry of confined magnetic textures using transferable hBN-based $V_B^-$ centers remains largely unexplored. In particular, magnetic microstructures based on permalloy (Py = $Ni_{81}Fe_{19}$), which host relevant vortex configurations and field-driven magnetization textures for spintronic applications, have not yet been quantitatively investigated using 2D $V_B^-$ quantum sensors[18–21]. A key challenge is to separate magnetic-field-induced ODMR shifts from non-magnetic variations of the resonance frequency, including local zero-field splitting and strain-induced inhomogeneity.

Here, we integrate an hBN flake containing $V_B^-$ centers with a 4 μm-diameter Py disk and perform spatially resolved ODMR magnetometry under ambient conditions. Correlative magnetic force microscopy (MFM) and ODMR measurements reveal a flux-closure vortex state at zero applied magnetic field and edge-localized magnetic surface charges under an in-plane magnetic field. By combining pixel-resolved baseline calibration with an edge-charge model, we reconstruct the out-of-plane stray-field distribution and relate the measured ODMR contrast to the field-driven vortex evolution. The proposed methodology enables quantitative extraction of magnetic-field distributions while discriminating magnetic contributions from local variations in the zero-field splitting, thereby extending the capabilities of 2D quantum sensors beyond qualitative magnetic imaging. Our results establish transferred hBN-based $V_B^-$ sensors as a practical platform for quantitative magnetometry of integrated magnetic microstructures.

## Results and Discussion

We first characterized the transferred hBN $V_B^-$ quantum sensor after integration with the Py microdisk. Figure 1(a) illustrates the hybrid quantum sensing platform and experimental configuration. $V_B^-$ centers were generated in a mechanically exfoliated hBN flake by 2.5 keV $He^+$ irradiation with a fluence of $3 \times 10^{14}$ $cm^{-2}$. The irradiated flake was subsequently transferred onto a 4 µm-diameter Py disk fabricated on a coplanar waveguide, placing the sensing layer in close proximity to the magnetic structure. Additional details of defect generation, device fabrication, and sample integration are provided in the Methods and Supporting Information Sections S1 and S2.

The sensing principle is based on the spin-triplet ($S = 1$) ground state of the $V_B^-$ center, shown schematically in Figure 1(b). At zero magnetic field, the $|0\rangle$ and $|\pm 1\rangle$ spin sublevels are separated by the zero-field splitting $D \approx 3.48\,GHz$. External magnetic fields $B_Z^{bias}$ lift the degeneracy between the $|-1\rangle$ and $|+1\rangle$ states through the Zeeman interaction, shifting the ODMR transition frequencies ($f_-$ and $f_+$) and enabling the local magnetic-field detection

Before investigating the magnetic microdisk, we first verified that the transferred $V_B^-$ ensembles had retained their characteristic optical and spin properties. The photoluminescence (PL) spectrum exhibits the broad emission band characteristic of $V_B^-$ centers between 700 and 900 nm (Figure 1(c). The ODMR spectra in Figure 1(d) exhibits the characteristic zero-field splitting and the expected Zeeman-induced separation of the spin transitions under an external magnetic field $B_Z^{bias}$ of 10 mT. These measurements confirm that the transferred hBN flake retains its spin-sensing functionality after device integration and remains suitable for quantitative magnetometry.

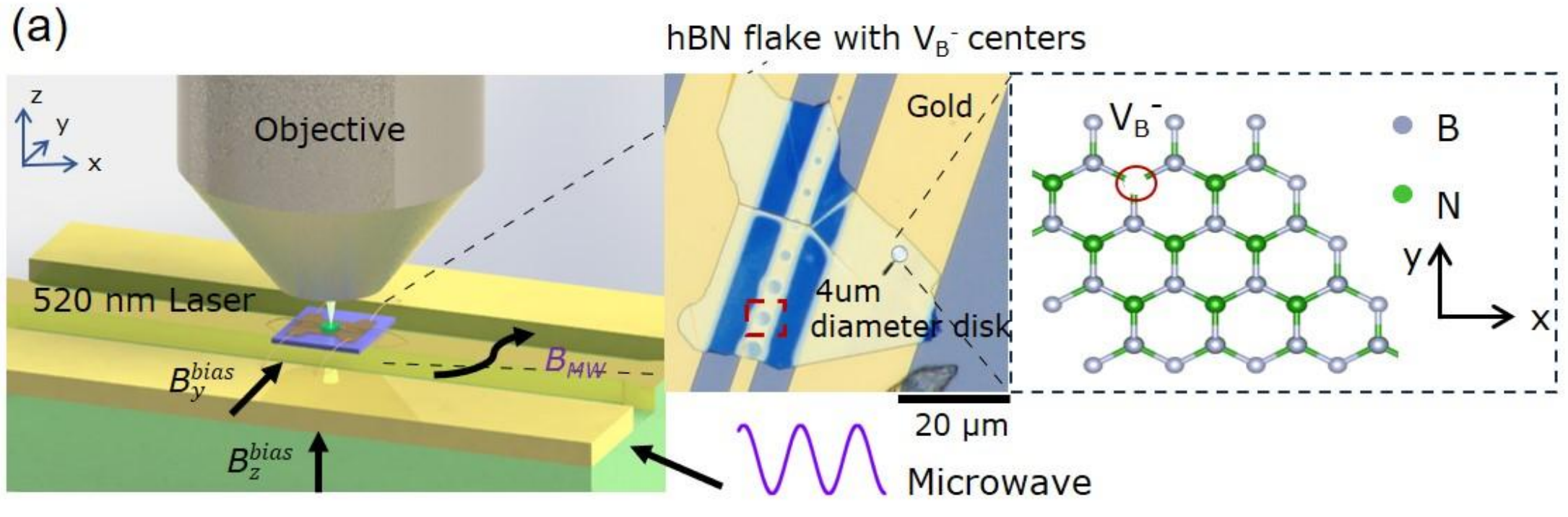

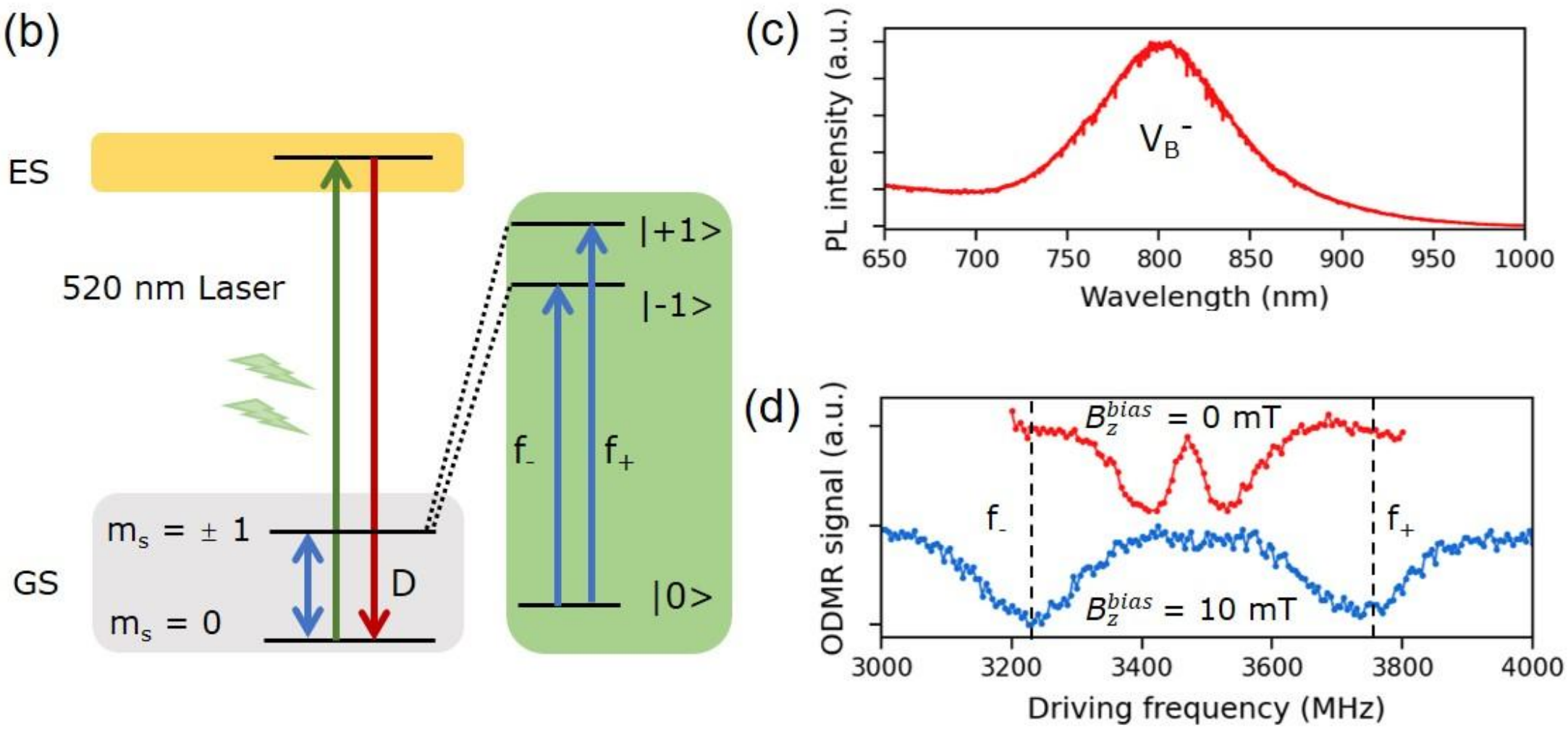

**Figure 1. Hybrid $V_B^-$ quantum sensing platform and sensor characterization.** (a) Schematic of the experimental configuration and optical microscope image of the transferred hBN flake integrated with a row of Py microdisks of different diameters fabricated on a coplanar waveguide; the present study focuses on the 4 μm-diameter disk. The right panel illustrates the atomic structure of the $V_B^-$ center. (b) Spin-dependent optical cycle and ground-state energy levels of the $V_B^-$ center. (c) Room-temperature photoluminescence spectrum of the $V_B^-$ centers in hBN. (d) ODMR spectra measured at zero magnetic field and under an external magnetic field $B_Z^{bias}$ of 10 mT, demonstrating the magnetic-field-dependent splitting of the spin transitions.

We next performed pixel-resolved ODMR mapping to extract the local spin-resonance parameters required for quantitative magnetometry. Quantitative magnetometry requires determining the spin-resonance frequencies at every spatial position while separating magnetic-field-induced frequency shifts from local variations in the intrinsic spin properties of the $V_B^-$ centers. In this protocol, a complete ODMR spectrum is recorded at each position of the scanned area, generating a three-dimensional dataset consisting of two spatial coordinates $(x, y)$ and the microwave frequency $(\nu)$, enabling quantitative analysis of the local spin-resonance parameters.

(Further details are provided in Supporting Information Section S3) Figure 2(a) shows the optical image of the investigated device together with the selected measurement region containing the 4 μm-diameter Py microdisk. The corresponding PL map is presented in Figure 2(b). The reduced PL intensity observed over part of the disk is attributed to non-uniform contact between the transferred hBN flake and the underlying Py surface, most likely caused by trapped interfacial gas pockets. Nevertheless, the PL intensity remains sufficient for reliable ODMR measurements.

A representative ODMR spectrum acquired from a single pixel is shown in Figure 2(c). This nonmagnetic reference region minimizes the influence of local magnetic-field gradients and allows reliable extraction of the intrinsic ODMR response.

To extract the local spin-resonance parameters, each ODMR spectrum was fitted using a double-Lorentzian function with a linear background[22,23]:

$$I(\nu) = c_0 + c_1\nu - \frac{A_-}{1+\frac{4(\nu-f_-)^2}{\Gamma_-^2}} - \frac{A_+}{1+\frac{4(\nu-f_+)^2}{\Gamma_+^2}} \quad (1)$$

where $I(\nu)$ is the ODMR signal as a function of microwave frequency $\nu$, $A_-$ and $A_+$ are the amplitudes of the two resonances, $\Gamma_-$ and $\Gamma_+$ are their full widths at half maximum, $f_-$ and $f_+$ are the fitted resonance frequencies, and $c_0$ and $c_1$ describe the linear background contribution. The effective spin parameters were then obtained as $D = (f_+ + f_-)/2$ and $E = (f_+ - f_-)/2$. Thus, $D$ and $E$ are derived from the fitted resonance frequencies rather than treated as independent fitting parameters. Spatial variations in $D$ primarily reflect changes in the local crystal environment, such as strain-induced shifts in the zero-field splitting, whereas $E$ contains the transverse splitting and magnetic-field-dependent information used for the quantitative magnetometry presented below[10,11]. The resulting spatial distributions of $D$ and $E$ at zero field are provided in Supporting Information Section S4.

The corresponding magnetic-field sensitivity, estimated from the ODMR linewidth, contrast, and photon count rate, is approximately 100 $\mu T/\sqrt{Hz}$, which is sufficient to resolve the stray fields generated by the Py disk. Further details are provided in Supporting Information Section S5.

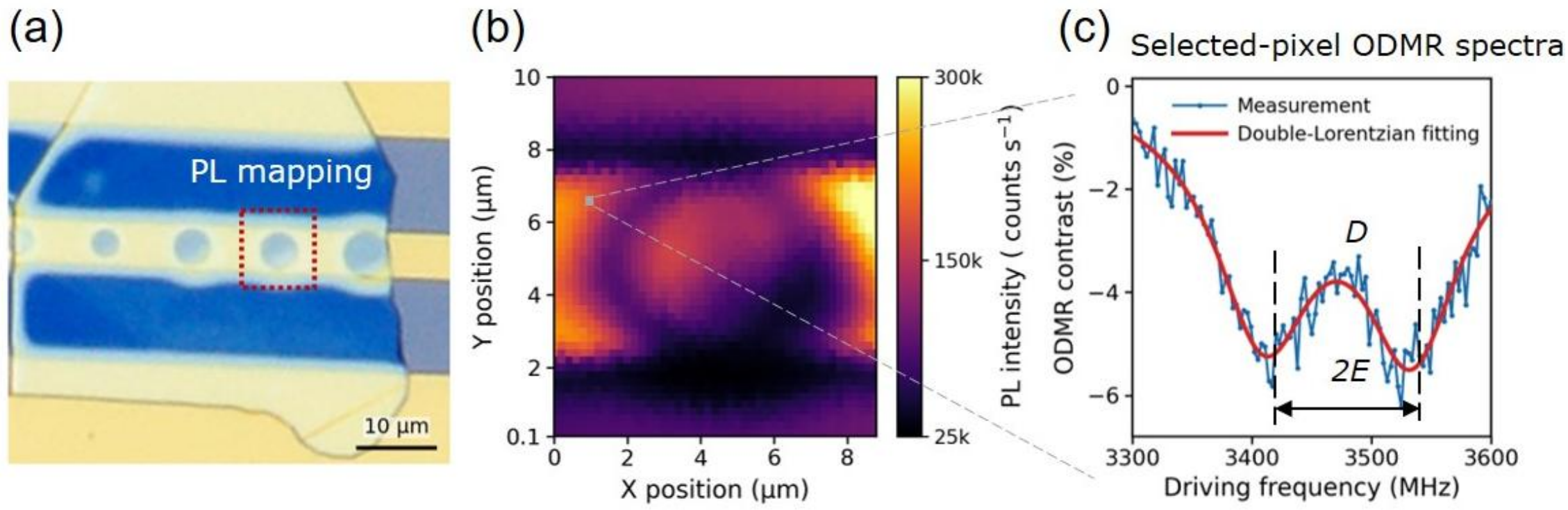

**Figure 2. Pixel-resolved ODMR analysis.** (a) Optical image of the investigated device showing the measurement region for the 4 μm-diameter Py microdisk (red dotted square). (b) PL map of the transferred $V_B^-$ ensembles within the selected area. (c) Representative ODMR spectrum recorded from a single selected pixel and the double-Lorentzian fitting used to extract the local spin-resonance parameters $D$ and $E$.

Having established the sensing response, we investigated the magnetic ground state of the Py microdisk and its evolution under an applied magnetic field using correlative MFM and ODMR measurements. Figure 3(a) shows the AFM topography of the 4 μm-diameter and the 50 nm-thick Py disk. The disk geometry is clearly resolved together with a raised rim and several surface particles originating from the fabrication process. These topographic features provide a structural reference for the subsequent magnetic characterization. The AFM image shows that the fabricated disk has a diameter of approximately 4.2 μm, in good agreement with the designed value of 4.0 μm, confirming reliable dimensional control during electron-beam lithography.

The corresponding zero-field MFM image in Figure 3(b) shows no extended bipolar or multidomain magnetic contrast. Instead, the disk interior exhibits a nearly uniform phase response, consistent with the formation of a flux-closure vortex state. In this configuration, the magnetization curls within the sample plane and remains approximately tangential to the disk boundary. The ideal surface magnetic-charge density at the disk sidewall is given by[25]

$$\sigma_m = \boldsymbol{M} \cdot \boldsymbol{n} \qquad (3)$$

where $\boldsymbol{M}$ is the magnetization and $\boldsymbol{n}$ is the outward surface normal. Because $\boldsymbol{M}$ is nearly perpendicular to $\boldsymbol{n}$ along most of the disk perimeter, and $\sigma_m$ approaches zero, strongly suppressing the stray magnetic field at the disk edge. The remaining out-of-plane magnetization is expected to be confined mainly to the vortex core, whose diameter is estimated to be approximately 16 nm using the Usov ansatz model (Supporting Information Section S6). Although a localized dark feature is visible in the MFM image, its position coincides with a surface particle observed in the AFM topography, indicating that it most likely originates from topographic cross-talk rather than from the magnetic vortex core. The absence of a clearly resolved vortex-core signal should nevertheless be interpreted with caution, because the MFM response is influenced by the finite tip response and the tip-sample separation. Overall, the nearly homogeneous MFM phase contrast, with the absence of extended multidomain contrast, supports the formation of a flux-closure vortex configuration in the Py disk[25,26].

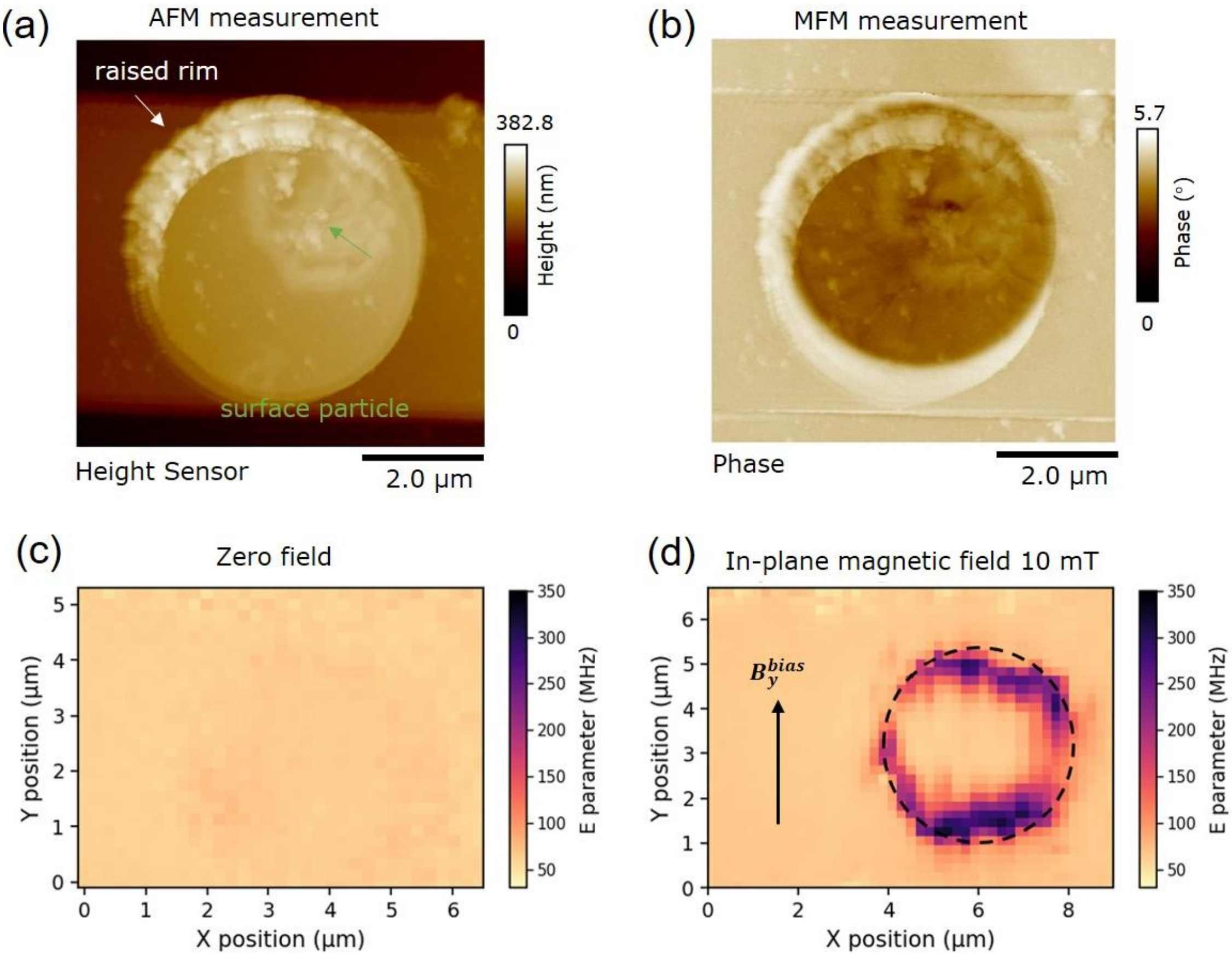

**Figure 3. Magnetic characterization of the vortex-state Py microdisk by MFM and ODMR.** (a) AFM topography of the fabricated Py microdisk on the coplanar waveguide, showing a measured disk diameter of approximately 4.2 μm, close to the nominal design diameter of 4.0 μm. (b) Corresponding zero-field MFM image showing no extended bipolar or multidomain magnetic

contrast, consistent with a flux-closure vortex state. (c) Zero-field ODMR-derived $E$-parameter map exhibiting a nearly homogeneous distribution dominated by intrinsic crystal-field variations of the $V_B^-$ ensembles. (d) ODMR-derived $E$-parameter map measured under an applied in-plane magnetic field $B_y^{bias}$ of 10 mT, directed predominantly along the $y$-axis of the mapping coordinate system. The enhanced ODMR splitting at opposite disk edges arises from magnetic surface charges generated by the field-driven distortion of the vortex state. The dashed circle indicates the disk boundary.

The corresponding ODMR-derived $E$-parameter map measured at zero applied field is shown in Figure 3(c). Using the same 0-350 MHz color scale as the field-on measurement, no pronounced enhancement is observed around the disk boundary, indicating that no spatially resolved Zeeman contribution from the Py stray field is detected under zero-field conditions. A narrower color scale, shown in Figure S7(b), reveals local variations of approximately 50-75 MHz, which are attributed predominantly to intrinsic transverse zero-field splitting and local crystal-field variations of the $V_B^-$ ensembles rather than to a pronounced magnetic contribution from the Py disk. These zero-field spatial variations are used as a local baseline in the subsequent magnetic-field reconstruction, allowing the field-induced contribution from the Py disk to be isolated. This behavior is consistent with the zero-field MFM measurements and confirms that the out of plane stray-field contribution from the Py disk is negligible under zero-field conditions. Moreover, because the estimated vortex-core diameter is far smaller than both the ODMR mapping step size (~200 nm) and the effective optical resolution of the microscope (~200 nm), the associated magnetic signal is spatially averaged and cannot be resolved in the present ODMR measurements.

Applying an in-plane magnetic field $B_y^{bias}$ of 10 mT fundamentally modifies the magnetic configuration of the disk. The resulting $E$-parameter map in Figure 3(d) shows two localized regions of enhanced splitting at opposite disk edges, while the disk interior and surrounding region remain weakly affected. This behavior reflects the field-driven displacement and distortion of the magnetic vortex. As the magnetization develops a finite component normal to the disk boundary, positive and negative magnetic surface charges are generated according to Equation (3). These magnetic surface charges produce pronounced stray-field lobes that are detected by the 2D hBN

$V_B^-$ sensor. The corresponding increase in the ODMR splitting is described by[10,11,16]

$$E_{eff} = \frac{f_+ - f_-}{2} \approx \sqrt{E_0^2 + (\gamma_e B_z)^2} \quad (4)$$

where $E_0$ denotes the intrinsic transverse zero-field splitting and $B_z$ is the out-of-plane stray-field component at the position of the hBN sensor, defined along the sample normal.

Because the ODMR splitting depends on the magnitude of the local magnetic field, magnetic surface charges of opposite polarity both appear as enhanced values in the $E$-parameter map. Although the $E$-parameter map clearly visualizes the field-induced redistribution of magnetic surface charges, it does not yet represent the quantitative magnetic-field distribution because the measured splitting still contains contributions from the intrinsic zero-field splitting of the ensemble $V_B^-$ centers. Separating these intrinsic contributions from the magnetic response is therefore essential for quantitative magnetometry and forms the basis of the magnetic-field reconstruction presented in the following section.

To convert the ODMR splitting into a quantitative magnetic-field distribution, we reconstructed the out-of-plane stray field after correcting for the local zero-field splitting and residual out-of-plane bias field. The latter arises because the permanent magnet was mounted on a rod, limiting precise angular control and introducing a weak out-of-plane field component. Quantitative reconstruction of the Py stray field therefore requires separating the local zero-field splitting and residual bias-field contributions from the measured ODMR response.

The reconstruction is based on the pixel-resolved ODMR splitting[10,11,16]

$$E_{eff}(x,y) = \frac{f_+(x,y) - f_-(x,y)}{2} = \sqrt{E_0^2(x,y) + (\gamma_e B_z(x,y))^2} \qquad (5)$$

where $E_0(x,y)$ denotes the local intrinsic transverse zero-field splitting extracted from the zero-field measurement, $\gamma_e$ is the electron gyromagnetic ratio, and $B_z(x,y)$ is the out-of-plane magnetic-field component at the position of the hBN sensor.

$$B_z(x,y) = B_z^{bias} + B_z^{stray}(x,y) \qquad (6)$$

where $B_z^{bias}$ is the residual out-of-plane bias field and $B_z^{stray}$ is the stray magnetic field generated by the Py disk. The local magnetic-field magnitude is therefore obtained from

$$|B_z(x,y)| = \frac{1}{\gamma_e}\sqrt{E_{eff}^2(x,y) - E_0^2(x,y)} \quad (7)$$

To improve numerical stability, the quantity $E_{eff}^2 - E_0^2$ was adaptively filtered before taking the square root, thereby suppressing noise amplification in regions where the magnetic field approaches zero. The residual out-of-plane bias field was estimated from the far-field ODMR response to be approximately 0.9 mT (The calculation is presented in Supporting Information Section S7) and subsequently removed from the reconstructed magnetic-field distribution.

The resulting out-of-plane stray-field map is shown in Figure 4(a). Two localized arc-shaped regions of enhanced magnetic field appear at opposite edges of the Py disk, whereas both the disk interior and the surrounding region exhibit only weak magnetic fields. The reconstructed field reaches a maximum magnitude of approximately 11.2 mT, demonstrating that the magnetic response is dominated by edge-localized magnetic surface charges. A circular fit to the reconstructed edge signal yields a disk at the center of (6.07, 2.96) μm and a radius of 1.94 μm, in good agreement with the device geometry.

Because the ODMR splitting depends only on the magnitude of the magnetic field, the inversion procedure does not directly determine the polarity of the reconstructed field. To assign the magnetic-field sign, we use the polarity predicted by the edge-charge model described in the following section. The resulting signed reconstruction is presented in Figure 4(b), revealing the expected dipolar field distribution with opposite magnetic polarities at the upper and lower edges of the disk. The reconstructed extrema reach approximately +11.2 mT and -10.2 mT. The known direction of the residual out-of-plane bias field provides the reference required to determine the global polarity of the reconstructed stray field map. These results show that pixel-resolved ODMR measurements from transferred hBN $V_B^-$ ensembles can be converted into quantitative magnetic-field maps, enabling a physically meaningful interpretation of the magnetic configuration.

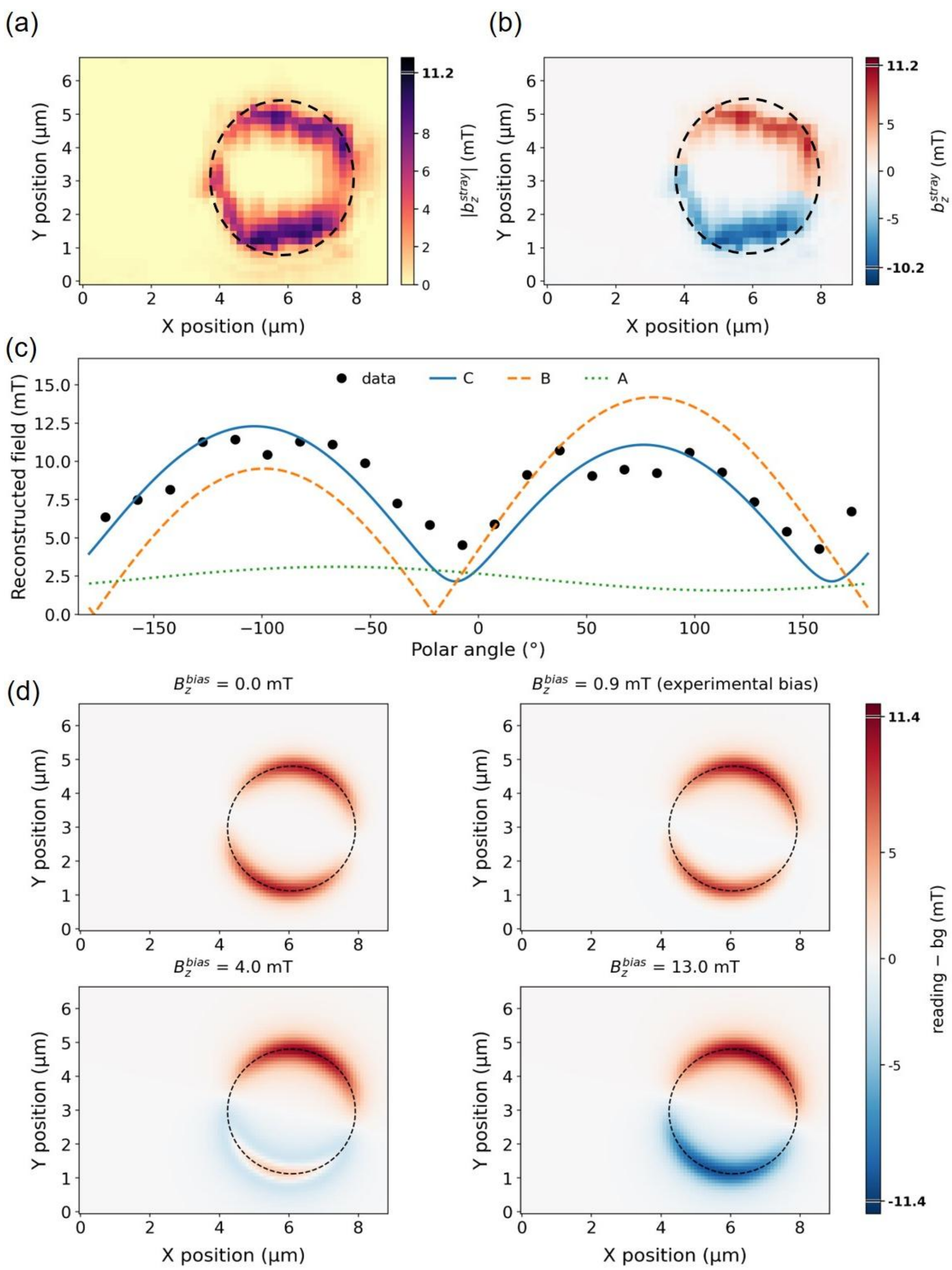

**Figure 4. Quantitative reconstruction and validation of the magnetic-field distribution in the Py microdisk.** (a) Reconstructed magnitude of the out-of-plane stray magnetic field $|B_z^{stray}|$, obtained by pixel-wise inversion of the ODMR splitting after correcting for the local zero-field splitting and residual out-of-plane bias field. The reconstructed field is localized at two opposite disk edges and reaches a maximum value of approximately 11.2 mT. (b) Model-assisted signed magnetic-field reconstruction obtained by assigning the local field polarity using the edge-charge model. The resulting dipolar field distribution exhibits opposite magnetic polarities at the upper and lower disk edges. (c) Angular magnetic-field profile extracted along the disk boundary and compared with three candidate fitting models. Model C provides the best agreement with the experimental data, yielding an RMS deviation of 1.67 mT, compared with 6.46 mT for Model A and 3.28 mT for Model B. The fitted peak field obtained from Model C, 11.5 mT, agrees with the reconstructed maximum field to within approximately 2%. (d) Forward-calculated ODMR response for residual out-of-plane bias fields of 0.0, 0.9, 4.0, and 13.0 mT. Increasing the bias field progressively transforms the inherently magnitude-sensitive ODMR response into a sign-sensitive dipolar magnetic-field distribution. Dashed circles indicate the fitted boundary of the Py microdisk.

To interpret the reconstructed magnetic-field distribution, we employ the edge-charge model describing the magnetostatic response of the distorted vortex state. Under an applied in-plane magnetic field, the magnetization develops a finite component normal to the disk boundary, giving rise to an effective magnetic surface-charge density[25].

$$\sigma_m(\theta) = M_{eff} \cos(\theta - \phi) \quad (8)$$

where $M_{eff}$ denotes the effective field-aligned magnetization and $\phi$ is its in-plane orientation. Positive and negative magnetic charges are therefore generated at opposite sides of the disk, producing the characteristic dipolar stray-field distribution.

To quantitatively evaluate the reconstruction, the angular field profile was extracted along an annular region surrounding the disk boundary and compared with three candidate readout models (Supporting Information Section S8). As shown in Figure 4(c), the best agreement is obtained using the rectified response Model C.

$$R_c(\theta) = \sqrt{[c + b_0 \cos(\theta - \phi)]^2 + (\frac{E_0}{\gamma_e})^2} \quad (9),$$

which accounts for both the intrinsic zero-field splitting and the residual out-of-plane bias field. Model C yields a root-mean-square deviation of only 1.67 mT, substantially lower than the alternative models. Furthermore, the fitted field amplitude (11.4 mT) agrees with the independently reconstructed peak field (11.2 mT) to within approximately 2%, demonstrating the internal consistency of the reconstruction procedure. The fitted angle, $\phi = 76°$ indicates that the field-aligned magnetization is oriented predominantly along the vertical (y-axis) direction of the mapping coordinate system. Owing to the magnitude-sensitive nature of ODMR in $V_B^-$ centers, this direction is determined modulo 180°.

The fitted edge profile further enables a semi-quantitative estimate of the effective field-aligned magnetization. Using the measured edge width together with the edge-charge model (Supporting Information Section S9), we estimate that the field-aligned magnetization is

$$\mu_0 M_{eff} \approx 0.51 \text{ T}$$

corresponding to approximately one-half of the saturation magnetization of permalloy. This result indicates that the applied magnetic field substantially distorts the vortex state but does not fully saturate the magnetic disk.

Finally, forward calculations were performed to evaluate the influence of the residual out-of-plane bias field on the ODMR response. The simulated maps are presented in Figure 4(d). In the absence of bias field, positive and negative edge magnetic surface charges produce identical ODMR splitting because the measurement is sensitive only to the magnetic-field magnitude. Increasing the out-of-plane bias progressively breaks this symmetry, causing the reconstructed response to evolve toward a sign-sensitive dipolar distribution. At the experimental bias field of approximately 0.9 mT, only a small asymmetry is introduced, whereas substantially larger bias fields would allow the magnetic-field polarity to be determined directly without model-assisted sign assignment. Further details of the forward calculations are provided in Supporting Information Section S10.

Overall, the agreement between the reconstructed magnetic-field distribution, the edge-charge model, and the forward ODMR calculations validates the pixel-resolved inversion approach for quantitative ODMR magnetometry of confined magnetic textures using a transferable 2D $V_B^-$ quantum sensor.

In summary, we have demonstrated spatially resolved quantitative magnetometry of a Py microdisk using a transferred hBN flake hosting an ensemble of $V_B^-$ centers. Pixel-resolved ODMR measurements, combined with local zero-field-splitting calibration and correction for a residual out-of-plane bias field, enable reconstruction of the out-of-plane stray magnetic field under ambient conditions. The reconstructed field reveals the field-driven distortion of the vortex-state magnetization and the emergence of edge-localized magnetic surface charges, with peak stray fields of approximately 11.2 mT. The agreement between the reconstructed magnetic-field distribution, the edge-charge model, and forward ODMR calculations validates the physical interpretation of the measured magnetic texture and establishes a quantitative ODMR inversion framework for transferred hBN quantum sensors. This approach provides a versatile platform for investigating magnetic nanostructures, spin-wave excitations, magnon transport, and hybrid quantum-spintronic systems.

## Methods

hBN flakes were mechanically exfoliated onto Si/$SiO_2$ substrates and implanted with 2.5 keV $He^+$ ions at a dose of $3 \times 10^{14}$ $cm^{-2}$ to generate $V_B^-$ centers. Ti/Au coplanar waveguides and Cr/Py/Cr magnetic microdisks were fabricated by electron-beam lithography and metal deposition, after which selected hBN flakes were transferred onto the Py disks using a PVA-based dry-transfer process. Spatially resolved ODMR measurements were performed using a custom-built confocal fluorescence microscope with 520 nm excitation, galvo scanning, photon counting with a gated Si avalanche single-photon detector, and microwave excitation delivered through the coplanar waveguide. MFM measurements were carried out in double-pass tapping-lift mode using magnetized low-moment tips. Additional experimental details are provided in the Supporting Information.

## Author Contributions

P.W., S.W., and Y.B. conceived and designed the experiments. Y.B. supervised the project, provided the overall scientific direction, and guided the interpretation of the results. K.W. and T.T. synthesized and provided the hBN single crystals. P.W., J.Q., Z.X., and K.S. fabricated the devices. P.W. performed the Raman and photoluminescence measurements. P.W. and R.S. carried out the AFM and MFM measurements. P.W. and S.W. performed the ODMR measurements; S.W. developed the ODMR measurement sequence and programmed the experimental control routines used for the spatially resolved ODMR experiments. P.W. developed and performed the edge-charge model simulations and related calculations, with input from Y.B. P.W., J.Q.,

K.S., S.P., S.Z., R.S., H.S., and Y.B. contributed to data visualization and interpretation. All authors analyzed the data and discussed the results. P.W. and Y.B. wrote the original draft of the manuscript, with input from all authors. All authors reviewed and edited the manuscript.

**Notes**

The authors declare no competing financial interest.

**Data Availability Statement**

The data supporting the findings of this study are available from the corresponding authors upon reasonable request.

**Supporting Information**

Additional details on $V_B^-$ defect creation and characterization, device fabrication, optical and microwave measurement setup, AFM imaging, spatial ODMR mapping, zero-field fitting parameters, vortex-core-size estimation, residual out-of-plane bias-field calculation, edge-charge model comparison, field-aligned magnetization estimation, and numerical stray-field reconstruction (PDF).

**Acknowledgments**

Support from the Ion Beam Center (IBC) and the Nanofabrication Facilities Rossendorf (NanoFaRo) at Helmholtz-Zentrum Dresden-Rossendorf (HZDR) is gratefully acknowledged. This work was partially funded by the German Federal Ministry of Research, Technology and Space (BMFTR; project no. 31 46229 007).